\documentclass[journal]{IEEEtran}
\IEEEoverridecommandlockouts
\usepackage{lineno}
\usepackage{cite}
\usepackage{hyperref}
\usepackage{amsmath,amssymb,amsfonts}
\usepackage{amsmath}
\usepackage{amsthm}

\usepackage{graphicx}
\usepackage{textcomp}
\usepackage{xcolor}
\usepackage{graphicx}
\usepackage{float}
\usepackage{subfigure}
\usepackage{amsmath}
\usepackage{amsfonts,amssymb}
\usepackage{mathrsfs}
\usepackage{mathtools}
\usepackage{algorithm}
\usepackage{algorithmicx}
\usepackage{algpseudocode}
\usepackage{bm}
\usepackage{multirow}
\usepackage{array}
\usepackage{amssymb}
\usepackage{amsmath}
\usepackage{cite}
\usepackage{url}
\usepackage{xcolor}
\usepackage{cite,graphicx,amsmath,amssymb}
\usepackage{subfigure}
\usepackage{fancyhdr}
\usepackage{mdwmath}
\usepackage{mdwtab}
\usepackage{caption}
\usepackage{amsthm}
\usepackage{setspace}
\usepackage{bm}
\usepackage{algorithm}
\usepackage{algpseudocode}
\usepackage{mathtools}
\usepackage{dsfont}
\usepackage{bbm}
\usepackage{yhmath}
\newtheorem{remark}{Remark}
\theoremstyle{definition}
\newtheorem{theorem}{Theorem}

\newtheorem{lemma}{Lemma}

\newtheorem{corollary}{Corollary}

\makeatletter
\newcommand{\biggg}{\bBigg@{3}}
\newcommand{\Biggg}{\bBigg@{3.5}}
\makeatother
\begin{document}
\title{Performance of Downlink and Uplink Integrated Sensing and Communications (ISAC) Systems}
\author{Chongjun~Ouyang, Yuanwei~Liu, and Hongwen~Yang
\thanks{C. Ouyang and H. Yang are with the School of Information and Communication Engineering, Beijing University of Posts and Telecommunications, Beijing, 100876, China (e-mail: \{DragonAim,yanghong\}@bupt.edu.cn).}
\thanks{Y. Liu is with the School of Electronic Engineering and Computer Science, Queen Mary University of London, London, E1 4NS, U.K. (e-mail: yuanwei.liu@qmul.ac.uk). (Corresponding author: Yuanwei Liu)}
}
\maketitle

\begin{abstract}
This letter analyzes the fundamental performance of integrated sensing and communications (ISAC) systems. For downlink and uplink ISAC, the diversity orders are analyzed to evaluate the communication rate (CR) and the high signal-to-noise ratio (SNR) slopes are unveiled for the CR as well as the sensing rate (SR). Furthermore, the achievable downlink and uplink CR-SR regions are characterized. It is shown that ISAC can provide more degrees of freedom for both the CR and the SR than conventional frequency-division sensing and communications systems where isolated frequency bands are used for sensing and communications, respectively.
\end{abstract}


\section{Introduction}
Enabling share of spectrum and hardware resources, integrated sensing and communications (ISAC) systems can perform dual-function sensing-communications within the same time-frequency resource block, which is expected to play a key role in the future wireless network market \cite{Wang2021,Liu2020,Mu2021,Liu2022}. Recently, ISAC has received considerable research attention due to its superior hardware- and spectral-efficiency compared to conventional frequency-division sensing and communications (FDSAC) systems where isolated frequency bands are used for sensing and communications, respectively \cite{Wang2021,Liu2020,Mu2021,Liu2022,Liu2021}.

Several works discussed the main features of ISAC and analyzed its performance from an information-theoretic perspective \cite{Chiriyath2016,Rong2018,Liu2021}. Typical information-theoretic performance metrics of ISAC include the estimation or sensing rate (SR) for radar sensing and the communication rate (CR) for communications. For more details about the performance of ISAC, please refer to the recent overview paper \cite{Liu2021} and references therein. Yet, it is worthy of mentioning that most previous ISAC researchers did not take account of the influence of channel fading when analyzing the performance of ISAC \cite{Chiriyath2016,Rong2018,Liu2021}. In addition, a rigorous discussion on the in-depth system insights of ISAC, including the diversity order and the high signal-to-noise ratio (SNR) slope, is still missing.

The aim of this letter is to analyze the performance of downlink and uplink ISAC from an information-theoretic perspective. To this aim, we discuss in detail the CR, SR, and achievable CR-SR region of ISAC by taking into account the capacity-achieving coding/decoding and the SR-optimal radar waveforming as well as the influence of channel fading. We further analyze the high-SNR CR and SR in order to unveil the diversity order and high-SNR slope. Theoretical analyses and numerical results indicate that ISAC is capable of providing more degrees of freedom for both the CR and the SR than conventional FDSAC.

\section{System Model}
\begin{figure}[!t]
    \centering
    \subfigbottomskip=0pt
	\subfigcapskip=-5pt
\setlength{\abovecaptionskip}{0pt}
   \subfigure[System model.]
    {
        \includegraphics[height=0.15\textwidth]{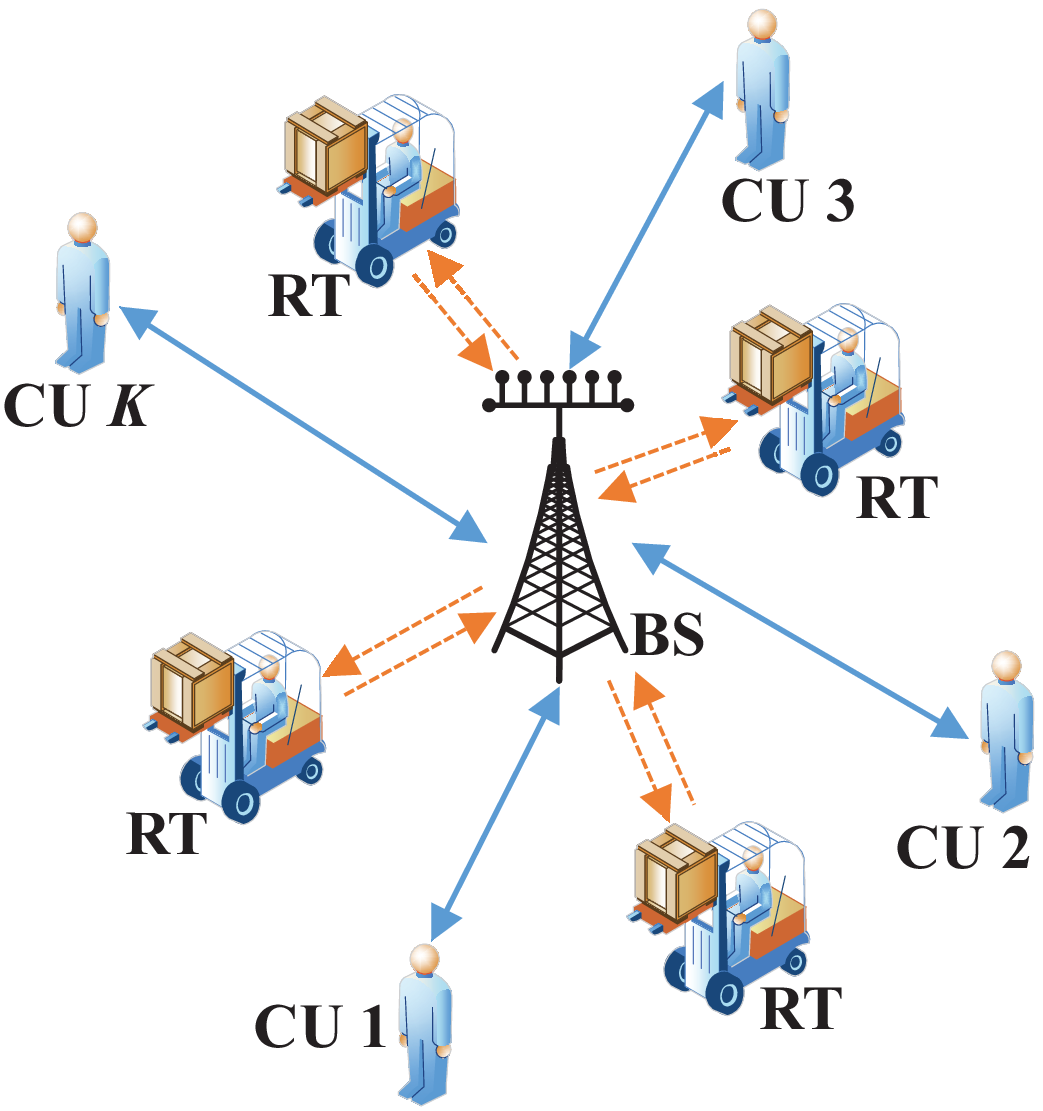}
	   \label{System_Model}	
    }
    \subfigure[RadCom BS.]
    {
        \includegraphics[height=0.15\textwidth]{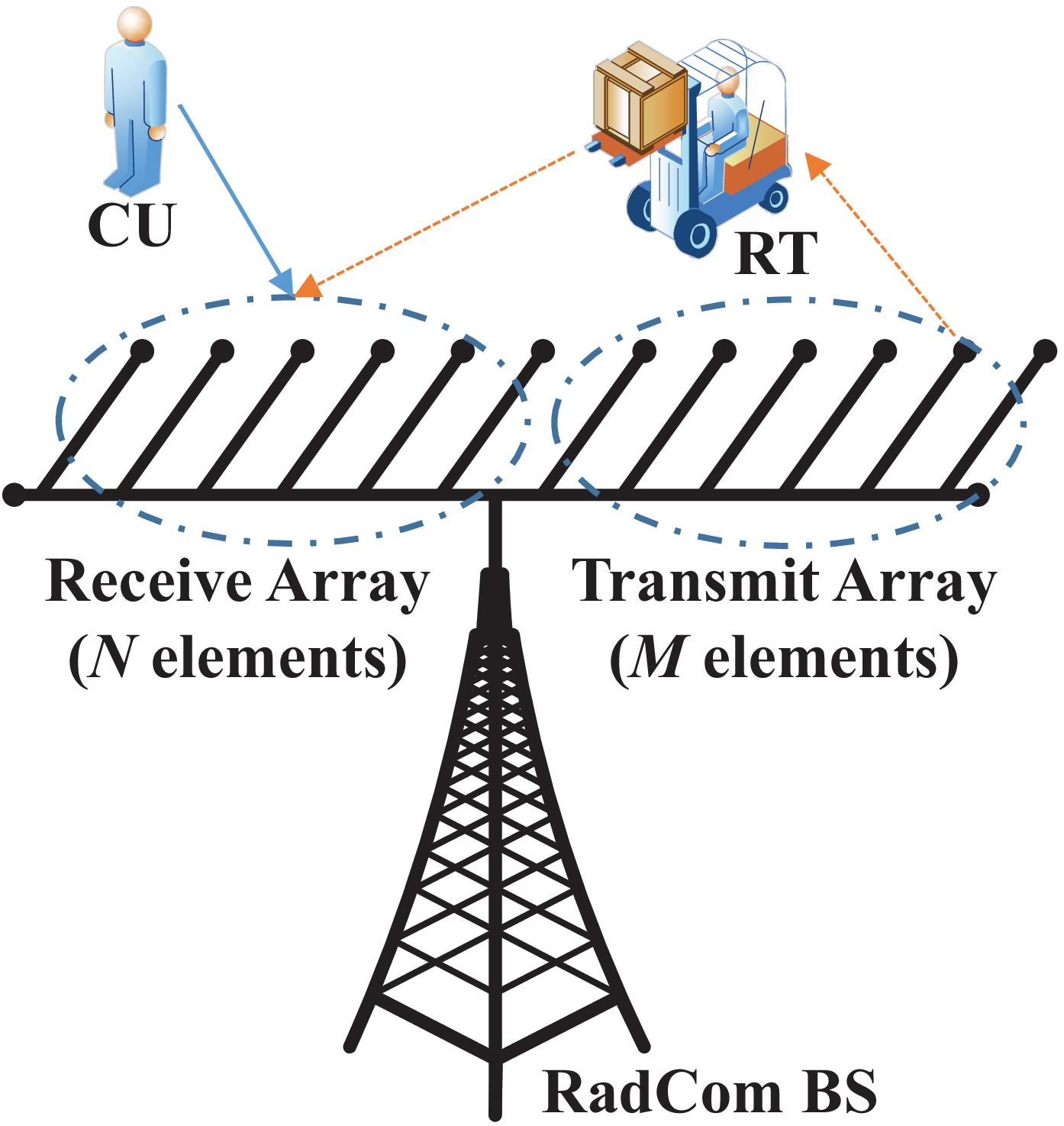}
	   \label{RadCom_BS}	
    }
\caption{An ISAC system with $K$ CUs and several RTs.}
    \label{figure1}
\end{figure}

In an ISAC system shown in {\figurename} \ref{System_Model}, one radar-communications (RadCom) base station (BS) serves $K$ single-antenna communication users (CUs) while simultaneously sensing the radar targets (RTs) in the near environment. The BS is equipped with two spatially widely separated antenna arrays, i.e., $M$ ($M\geq K$) transmit antennas and $N$ ($N\geq K$) receive antennas, whose structure is illustrated in {\figurename} \ref{RadCom_BS}. In this letter, a structure of statistical multiple-input multiple-output (MIMO) radar is considered, where the receive antennas are widely separated \cite{Tang2019}. In this case, we can ignore the spatial correlation between receive antennas \cite{Tang2019}.

\subsection{Downlink ISAC}
The downlink ISAC (D-ISAC) comprises two stages. In the first stage, the BS broadcasts the communication signal plus radar waveform to the CUs and RTs. Accordingly, the received signal at CU $k\in{\mathcal{K}}=\left\{1,\cdots,K\right\}$ is given by
\begin{align}
{{\textbf{y}}}_{{\text{d}},k}^{\mathsf{H}}={\textbf{h}}_{{\text{d}},k}^{\mathsf{H}}\left({\textbf{X}}_{\text{d}}+{\textbf{S}}\right)+{\textbf{n}}_{{\text{d}},k}^{\mathsf{H}},
\end{align}
where ${{\textbf{y}}}_{{\text{d}},k}\in{\mathbbmss{C}}^{L\times 1}$ with the subscript ``${\text{d}}$'' denoting downlink transmission; ${\textbf{h}}_{{\text{d}},k}^{\mathsf{H}}\in{\mathbbmss{C}}^{1\times M}$ is the downlink channel vector from the transmit array at the BS to CU $k$; ${\textbf{n}}_{{\text{d}},k}\sim{\mathcal{CN}}\left({\textbf{0}},{\textbf{I}}_L\right)$ is the additive white Gaussian noise (AWGN); ${\textbf{S}}=\left[{\textbf{s}}_1 \cdots {\textbf{s}}_L\right]\in{\mathbbmss{C}}^{M\times L}$ ($L\geq M$, $L\geq N$) is the radar waveform with ${\textbf{s}}_l\in{\mathbbmss C}^{M\times1}$ ($l\in\mathcal{L}=\left\{1,\cdots,L\right\}$) representing the waveform at the $l$th time slot; ${\textbf{X}}_{\text{d}}=\left[{\textbf{x}}_{{\text{d}},1} \cdots {\textbf{x}}_{{\text{d}},L}\right]\in{\mathbbmss{C}}^{M\times L}$ is the communication signal matrix with ${\textbf{x}}_{{\text{d}},l}\in{\mathbbmss C}^{M\times1}$ representing the downlink communication signal at the $l$th time slot. Moreover, the communication signal and the radar waveform are subject to the power budget ${\mathbbmss{E}}\{{\textbf{x}}_{{\text{d}},l}^{\mathsf{H}}{\textbf{x}}_{{\text{d}},l}\}\leq p_{\text{c}}$ ($\forall l\in\mathcal{L}$) and ${\mathsf{tr}}({\textbf{S}}{\textbf{S}}^{\mathsf{H}})\leq p_{\text{s}}$, respectively, where $p_{\text{c}}$ and $p_{\text{s}}$ denote the communication SNR and sensing SNR, respectively. Generally, $p_{\text{c}}$ and $p_{\text{s}}$ should be subject to a sum power constraint. Yet, in this letter, for brevity, we consider $p_{\text{s}}$ and $p_{\text{c}}$ are fixed values, which can be treated as a case of fixed power allocation \cite{Chiriyath2016,Rong2018}. The influence of power allocation will be discussed in our future works. In the second stage of the downlink ISAC, the BS aims to extract environmental information from the reflected radar echoes. Particularly, the signal received by the BS is given by ${\textbf{Y}}={\textbf{G}}^{\mathsf{H}}{\textbf{X}}_{\text{s}}+{\textbf{G}}_{\text{c}}^{\mathsf{H}}{\textbf{X}}_{\text{s}}+{\textbf{N}}^{\mathsf{H}}$, where ${\textbf{X}}_{\text{s}}={\textbf{S}}+{\textbf{X}}_{{\text{d}}}\in{\mathbbmss C}^{M\times L}$; $\textbf{G}=\left[{\textbf{g}}_1\cdots{\textbf{g}}_N\right]\in{\mathbbmss{C}}^{M\times N}$ is the target response matrix (TRM) of the RTs; ${\textbf{N}}=\left[{\textbf{n}}_1 \cdots {\textbf{n}}_N\right]\in{\mathbbmss{C}}^{L\times N}$ is the AWGN; and $\textbf{G}_{\text{c}}\in{\mathbbmss{C}}^{M\times N}$ is the TRM of the CUs. Since all the CUs are registered users in the system, it makes sense to assume that $\textbf{G}_{\text{c}}$ is estimated in advance with conventional estimation algorithms and the signals reflected by the CUs, i.e., ${\textbf{G}}_{\text{c}}^{\mathsf{H}}{\textbf{X}}_{\text{s}}$, are perfectly removed at the BS \cite{Wang2021,Mu2021,Liu2022}. Thus, the following signal is exploited to perform radar target sensing:
\begin{align}\label{Downlink_Sensing_Model}
{\textbf{Y}}_{{\text{d}}}={\textbf{G}}^{\mathsf{H}}\left({\textbf{S}}+{\textbf{X}}_{{\text{d}}}\right)+{\textbf{N}}^{\mathsf{H}}\in{\mathbbmss{C}}^{N\times L}.
\end{align}
Particularly, we assume that ${\textbf{h}}_{{\text{d}},k}\sim{\mathcal{CN}}\left({\textbf{0}},{\textbf{R}}_k\right)$ with ${\textbf{R}}_k\in{\mathbbmss{C}}^{M\times M}$ denoting the transmit correlation matrix and ${\mathbbmss{E}}\{{\textbf{h}}_{{\text{d}},k}{\textbf{h}}_{{\text{d}},k'}^{\mathsf{H}}\}={\textbf{0}}$ ($\forall k'\neq k$). Besides, we have ${\textbf{g}}_n\sim{\mathcal{CN}}\left({\textbf{0}},{{\textbf{R}}}_{\text{T}}\right)$ ($\forall n\in{\mathcal{N}}=\left\{1,\cdots,N\right\}$) with ${\textbf{R}}_{\text{T}}\in{\mathbbmss{C}}^{M\times M}$ being the transmit correlation matrix, ${\mathbbmss{E}}\{{\textbf{g}}_n{\textbf{g}}_{n'}^{\mathsf{H}}\}={\textbf{0}}$ ($\forall n\neq n'$), ${\textbf{n}}_n\sim{\mathcal{CN}}\left({\textbf{0}},{\textbf{I}}_L\right)$ ($\forall n\in{\mathcal{N}}$), and ${\mathbbmss{E}}\{{\textbf{n}}_n{\textbf{n}}_{n'}^{\mathsf{H}}\}={\textbf{0}}$ ($\forall n\neq n'$). Throughout this paper, full channel state information (CSI) of CU $k$ ($\forall k\in{\mathcal{K}}$) is assumed to be known to the BS and CU $k$ for the sake of discussing the performance upper bound of communications \cite{Mu2021,Wang2021,Liu2022}. Moreover, it is widely known that the TRM $\textbf{G}$ contains all the information of the RTs, such as the direction of each RT, and thus the radar target sensing can be regarded as the estimation of $\textbf{G}$ \cite{Liu2022,Tang2019,Yang2007}. Since $\textbf{G}$ needs to be sensed, the BS is assumed to know only the spatial correlation matrix, ${\textbf{R}}_{\text{T}}$. For the sake of brevity, we consider the case of ${\textbf{R}}_{\text{T}}\succ{\textbf{0}}$ throughout this letter.

\subsection{Uplink ISAC}
The uplink ISAC (U-ISAC) also includes two stages. Firstly, the BS broadcasts the radar waveform ${\textbf{S}}$ for sensing the nearby environment. Secondly, the BS receives the radar waveform reflected by the RTs and the communication messages sent by the CUs simultaneously. The communication signals and sensing signals are assumed to be synchronized perfectly at the BS by the method in \cite{Liu2020}. Besides, like the downlink case, we assume the BS can remove the radar echoes reflected by the CUs. Thus, the BS can decode the communication data as well as sensing the radar target from the signal as follows:
\begin{align}\label{Uplink_Basic_Model}
{\textbf{Y}}_{{\text{u}}}=\sum\nolimits_{k=1}^{K}{{\textbf{h}}}_{{\text{u}},k}{\textbf{x}}_{{\text{u}},k}^{\mathsf{H}}+
{\textbf{G}}^{\mathsf{H}}{\textbf{S}}+{\textbf{N}}^{\mathsf{H}},
\end{align}
where ${\textbf{Y}}_{{\text{u}}}=\left[{\textbf{y}}_{{\text{u}},1}\cdots{\textbf{y}}_{{\text{u}},L}\right]\in{\mathbbmss{C}}^{N\times L}$ with the subscript ``${\text{u}}$'' denoting uplink transmission; ${\textbf{h}}_{{\text{u}},k}\in{\mathbbmss{C}}^{N\times1}$ is the uplink channel vector from CU $k$ to the receive antenna array of the BS; ${\textbf{x}}_{{\text{u}},k}=\left[x_{{\text{u}},k,1},\cdots,x_{{\text{u}},k,L}\right]^{\mathsf{H}}\in{\mathbbmss{C}}^{L\times1}$ is the message sent by CU $k$ subject to the power budget ${\mathbbmss{E}}\{\left|x_{{\text{u}},k,l}\right|^2\}\leq p_{\text{c}}$ ($\forall l\in{\mathcal{L}}$). As explained earlier, the correlation between receive antennas can be omitted, and thus we can assume ${\textbf{h}}_{{\text{u}},k}\sim{\mathcal{CN}}\left({\textbf{0}},{\textbf{I}}_N\right)$ and ${\mathbbmss{E}}\{{\textbf{h}}_{{\text{u}},k}{\textbf{h}}_{{\text{u}},k'}^{\mathsf{H}}\}={\textbf{0}}$ ($\forall k\neq k'$). Besides, we assume the BS knows the full information of ${\textbf{h}}_{{\text{u}},k}$ ($\forall k\in\mathcal{K}$) and ${\textbf{R}}_{\text{T}}$. After the BS receives ${\textbf{Y}}_{\text{u}}$ presented in \eqref{Uplink_Basic_Model}, it can leverage a successive interference cancellation (SIC)-based framework to decode the communication signal, ${\textbf{x}}_{{\text{u}},k}$, as well as sensing the TRM, $\textbf{G}$ \cite{Chiriyath2016}. Specifically, the BS first decodes ${\textbf{x}}_{{\text{u}},k}$ by treating the radar waveform as interference. Then, ${\textbf{x}}_{{\text{u}},k}$ can be subtracted from ${\textbf{Y}}_{\text{u}}$ and the rest part will be used for sensing.

\section{Downlink Performance}
\subsection{Performance of Communications}
In this letter, we assume only the statistical information of the communication signal is used during the design of the radar waveform. Thus, CU $k$ can know the designed waveform in advance and remove the term ${\textbf{h}}_{{\text{d}},k}^{\mathsf{H}}{\textbf{S}}$ from ${{\textbf{y}}}_{{\text{d}},k}^{\mathsf{H}}$ before decoding the information bits. Besides, to analyze the performance upper bound of communication signals, we exploit dirty paper coding (DPC) to generate ${\textbf{X}}_{\text{d}}$, which can achieve the sum CR capacity of broadcast channels. Under the uplink-downlink duality, the maximal downlink sum CR is given by \cite{Heath2018}
\begin{align}
{\mathcal{R}}_{{\text{d}}}=\max_{\sum_{k=1}^{K}p_k\leq p_{\text{c}}}\log_2\det\left({\textbf{I}}_M+\sum\nolimits_{k=1}^{K}p_k{\textbf{h}}_{{\text{d}},k}{\textbf{h}}_{{\text{d}},k}^{\mathsf{H}}\right).
\end{align}
\subsubsection{Outage Probability}
The outage probability (OP) of the sum downlink CR is given by $P_{{\text{d}}}=\Pr\left({\mathcal{R}}_{{\text{d}}}<{\mathcal{R}}\right)$, where ${\mathcal{R}}$ denotes the target rate. Yet, ${\mathcal{R}}_{\text{d}}$ lacks any closed-form solutions, which together with the fact that $\left\{{\textbf{h}}_{{\text{d}},k}\right\}_{k=1}^{K}$ are independent but not identically distributed random vectors, makes the quantitative analysis of $P_{{\text{d}}}$ an intractable problem. As a compromise, we assume all the CUs share the same correlation matrix to glean further insights. In this case, the following theorem can be found.
\vspace{-5pt}
\begin{theorem}\label{Theorem_Downlink_OP_Asymptotic}
When ${\textbf{R}}_k={\textbf{R}}$ ($\forall k\in{\mathcal{K}}$), the OP satisfies
\begin{align}
\lim\nolimits_{p_{\text{c}}\rightarrow\infty}P_{{\text{d}}}= {\mathcal{O}}\left(p_{\text{c}}^{-MK}\right).
\end{align}
\end{theorem}
\vspace{-5pt}
\begin{IEEEproof}
Please refer to Appendix \ref{Proof_Theorem_Downlink_OP_Asymptotic} for more details.
\end{IEEEproof}
\vspace{-5pt}
\begin{remark}
When ${\textbf{R}}_k={\textbf{R}}$ ($\forall k$), a diversity order of $KM$ is achievable for the sum communication rate of the CUs.
\end{remark}
\vspace{-5pt}
\subsubsection{Ergodic Rate}
The ergodic CR (ECR) of the CUs in downlink transmission is given by ${\mathcal{R}}_{{\text{d}},{\text{c}}}={\mathbbmss{E}}\left\{{\mathcal{R}}_{{\text{d}}}\right\}$. Let ${\textbf{H}}_{\text{d}}=\left[{\textbf{h}}_{{\text{d}},1}\cdots{\textbf{h}}_{{\text{d}},K}\right]\in{\mathbbmss{C}}^{M\times K}$ denote the concatenation of the channels. Then, the following theorem can be found.
\vspace{-5pt}
\begin{theorem}\label{Theorem_Downlink_ECR_Asymptotic}
The downlink ECR satisfies
\begin{align}
\lim_{p_{\text{c}}\rightarrow\infty}{\mathcal{R}}_{{\text{d}},{\text{c}}}= K\log_2{\frac{p_{\text{c}}}{K}}+{\mathbbmss{E}}\left\{\log_2\det\left({\textbf{H}}_{\text{d}}^{\mathsf{H}}{\textbf{H}}_{\text{d}}\right)\right\}.
\end{align}
\end{theorem}
\vspace{-5pt}
\begin{IEEEproof}
Please refer to Appendix \ref{Proof_Theorem_Downlink_ECR_Asymptotic} for more details.
\end{IEEEproof}
Note that $\mathcal{E}_{\text{d}}={\mathbbmss{E}}\left\{\log_2\det\left({\textbf{H}}_{\text{d}}^{\mathsf{H}}{\textbf{H}}_{\text{d}}\right)\right\}$ is a constant independent of $p_{\text{c}}$, which lacks any closed-form expressions. Yet, when ${\textbf{R}}_k={\textbf{I}}_M$ ($\forall k\in{\mathcal{K}}$), $\mathcal{E}_{\text{d}}$ can be calculated as follows.
\vspace{-5pt}
\begin{corollary}\label{corollary1}
When ${\textbf{R}}_k={\textbf{I}}_M$ ($\forall k\in{\mathcal{K}}$), we have $\mathcal{E}_{\text{d}}=\frac{1}{\ln{2}}\sum\nolimits_{t=0}^{K-1}\left(
\sum\nolimits_{a=1}^{M-t-1}\frac{1}{a}-{\emph{\textbf{C}}}\right)$, where ${\emph{\textbf{C}}}$ is the Euler constant.
\end{corollary}
\vspace{-5pt}
\begin{IEEEproof}
Please refer to Appendix \ref{Proof_Theorem_Downlink_ECR_Asymptotic} for more details.
\end{IEEEproof}
\vspace{-5pt}
\begin{remark}
A high-SNR slope of $K$ is achievable for the downlink sum communication rate.
\end{remark}
\vspace{-5pt}

\subsection{Performance of Sensing}
Turn now to the sensing performance. The BS can use the received signal presented in \eqref{Downlink_Sensing_Model} to sense the TRM $\textbf{G}$. The performance of radar target sensing is evaluated by the SR that is defined as the sensing mutual information (MI) per unit time \cite{Tang2019}. In particular, the sensing MI is the MI between the received signal ${\textbf{Y}}_{\text{d}}$ and the TRM ${\textbf{G}}$ for a given $\textbf{S}$ \cite{Tang2019}. There are two reasons for using the SR as the sensing performance metric. The first reason is that the SR tells how much environmental information can be extracted from ${\textbf{Y}}_{\text{d}}$ with the view of information theory. The second reason is that under our considered ISAC model where ${\textbf{g}}_n\sim{\mathcal{CN}}\left({\textbf{0}},{{\textbf{R}}}_{\text{T}}\right)$, ${\textbf{n}}_n\sim{\mathcal{CN}}\left({\textbf{0}},{\textbf{I}}_L\right)$ ($\forall n\in{\mathcal{N}}$), ${\mathbbmss{E}}\left\{{\textbf{g}}_n{\textbf{g}}_{n'}^{\mathsf{H}}\right\}={\textbf{0}}$, ${\mathbbmss{E}}\left\{{\textbf{n}}_n{\textbf{n}}_{n'}^{\mathsf{H}}\right\}={\textbf{0}}$ ($\forall n\neq n'$), and ${\textbf{R}}_{\text{T}}\succ{\textbf{0}}$, the optimal radar waveform based on maximizing the SR has the same estimation performance as that based on minimizing the mean-square error (MSE) in estimating the TRM $\textbf{G}$ \cite{Yang2007}. For more details about the relationship between the SR (or the sensing MI) and the MSE, please refer to \cite{Yang2007}. In this letter, we assume that each waveform symbol lasts 1 unit time. Thus, the SR can be calculated as ${\mathcal{I}}_{{\text{d}},L}/L$, where ${\mathcal{I}}_{{\text{d}},L}$ denotes the sensing MI over the duration of $L$ symbols. By definition, we have ${\mathcal{I}}_{{\text{d}},L}=I\left({\textbf{Y}}_{\text{d}};{\textbf{G}}|{\textbf{S}}\right)$, where $I\left(X;Y|Z\right)$ denotes the MI between $X$ and $Y$ conditioned on $Z$. To simplify the expression of ${\mathcal{I}}_{{\text{d}},L}$ as well as the subsequent analyses, we treat ${\textbf{G}}^{\mathsf{H}}{\textbf{X}}_{\text{d}}$ as interference, which thus yields a sensing performance lower bound. Moreover, from a worst-case design perspective \cite{Hassibi2003}, the aggregate interference-plus-noise ${\textbf{Z}}={\textbf{G}}^{\mathsf{H}}{\textbf{X}}_{\text{d}}+{\textbf{N}}\in{\mathbbmss{C}}^{N\times L}$ is treated as the Gaussian noise. Since ${\textbf{X}}_{\text{d}}$ is dirty paper coded, we have ${\mathbbmss{E}}\{{\textbf{x}}_{{\text{d}},l}{\textbf{x}}_{{\text{d}},l'}^{\mathsf{H}}\}={\textbf{0}}$ ($\forall l\neq l'$) and ${\textbf{x}}_{{\text{d}},l}\sim{\mathcal{CN}}\left({\textbf{0}}_M,{\bm\Sigma}_{{\textbf{H}}_{\text{d}}}\right)$ ($\forall l\in{\mathcal{L}}$) with ${\mathsf{tr}}\left({\bm\Sigma}_{{{\textbf{H}}_{\text{d}}}}\right)\leq p_{\text{c}}$, where ${\bm\Sigma}_{{\textbf{H}}_{\text{d}}}$ is obtained by the iterative water-filling method \cite{Heath2018}. On this basis, we characterize the sensing MI as follows.
\vspace{-5pt}
\begin{lemma}\label{Lemma_Downlink_Sensing_MI}
The sensing MI can be written as ${\mathcal{I}}_{{\text{d}},L}=N\log_2\det({\textbf{I}}_L+{\sigma^{-2}}{\textbf{S}}^{\mathsf{H}}{\textbf{R}}_{\text{T}}{\textbf{S}})$
with $\sigma^2=1+{\mathsf{tr}}\left({\textbf{R}}_{\text{T}}{\bm\Sigma}\right)$ and ${\bm\Sigma}={\mathbbmss{E}}_{{\textbf{H}}_{\text{d}}}\left\{{\bm\Sigma}_{{\textbf{H}}_{\text{d}}}\right\}\leq p_{\text{c}}$.
\end{lemma}
\vspace{-5pt}
\begin{IEEEproof}
Please refer to Appendix \ref{Proof_Lemma_Downlink_Sensing_MI} for more details.
\end{IEEEproof}
We comment that ${\bm\Sigma}$ lacks any closed-form expressions, which can be evaluated numerically. Based on Lemma \ref{Lemma_Downlink_Sensing_MI}, the maximal downlink SR can be expressed as ${\mathcal{R}}_{{\text{d}},{\text{s}}}=\frac{1}{L}\max\nolimits_{{\mathsf{tr}}\left({\textbf{S}}{\textbf{S}}^{\mathsf{H}}\right)\leq p_{\text{s}}}{\mathcal{I}}_{{\text{d}},L}$. Theorem \ref{Theorem_Downlink_SR} provides an exact expression for ${\mathcal{R}}_{{\text{d}},{\text{s}}}$ as well as its high-SNR approximation.
\vspace{-5pt}
\begin{theorem}\label{Theorem_Downlink_SR}
The maximal downlink SR is given by
\begin{align}
{\mathcal{R}}_{{\text{d}},{\text{s}}}={N}{L^{-1}}\sum\nolimits_{m=1}^{M}\log_2\left(1+{\sigma^{-2}}\lambda_ms_m^{\star}\right),
\end{align}
where $\left\{\lambda_m>0\right\}_{m=1}^{M}$ denote the eigenvalues of ${\textbf{R}}_{\text{T}}$ and $s_{m}^{\star}=\max\left\{0,\frac{1}{\nu}-\frac{\sigma^2}{\lambda_m}\right\}$ with $\sum_{m=1}^{M}\max\left\{0,\frac{1}{\nu}-\frac{\sigma^2}{\lambda_m}\right\}=p_{\text{s}}$. The maximal SR is achieved when ${\textbf{S}}{\textbf{S}}^{\mathsf{H}}={\textbf{U}}_{\text{T}}^{\mathsf{H}}{\bm\Delta}^{\star}{\textbf{U}}_{\text{T}}$, where ${\textbf{U}}_{\text{T}}^{\mathsf{H}}{\mathsf{diag}}\left\{\lambda_1,\cdots,\lambda_{M}\right\}{\textbf{U}}_{\text{T}}$ denotes the eigendecomposition (ED) of ${\textbf{R}}_{\text{T}}$ and ${\bm\Delta}^{\star}={\mathsf{diag}}\left\{s_1^{\star},\cdots,s_M^{\star}\right\}$.
When $p_{\text{s}}\rightarrow\infty$, we can obtain
\begin{align}
{\mathcal{R}}_{{\text{d}},{\text{s}}}
\approx\frac{NM}{L}\left(\log_2{p_{\text{s}}}+\frac{1}{M}\sum\nolimits_{m=1}^{M}\log_2\left(\frac{\lambda_m}{M\sigma^2}\right)\right).
\end{align}
\end{theorem}
\vspace{-5pt}
\begin{IEEEproof}
Please refer to Appendix \ref{Proof_Theorem_Downlink_SR} for more details.
\end{IEEEproof}
\vspace{-5pt}
\begin{remark}
A high-SNR slope of $\frac{NM}{L}$ is achievable for the maximal downlink SR.
\end{remark}
\vspace{-5pt}
\subsection{Performance of FDSAC}\label{Section3c}
Turn now to the performance of downlink FDSAC (D-FDSAC) systems, where the total bandwidth is separated into two sub-bands, one for sensing only and the other for communications. It is assumed that $\alpha\in\left[0,1\right]$ fraction of the total bandwidth is used for communications. Besides, the DPC and the optimal radar waveforming \cite{Tang2019} are exploited to generate communication and sensing signals, respectively. In this case, the ECR is given by ${\mathcal{R}}_{{\text{d}},{\text{c}}}^{\alpha}={\mathbbmss{E}}\left\{\alpha\max_{\sum_{k=1}^{K}p_i\leq p_{\text{c}}}\log_2\det\left({\textbf{I}}_M+\sum\nolimits_{k=1}^{K}\frac{p_k}{\alpha}{\textbf{h}}_{{\text{d}},k}{\textbf{h}}_{{\text{d}},k}^{\mathsf{H}}\right)\right\}$. As for radar sensing, the maximal downlink SR is ${\mathcal{R}}_{{\text{d}},{\text{s}}}^{\alpha}=\frac{N(1-\alpha)}{L}\max_{{\mathsf{tr}}\left({\textbf{S}}{\textbf{S}}^{\mathsf{H}}\right)\leq p_{\text{s}}}\log_2\det({\textbf{I}}_L+\frac{1}{1-\alpha}{\textbf{S}}^{\mathsf{H}}{\textbf{R}}_{\text{T}}{\textbf{S}})$. It is worth noting that $({\mathcal{R}}_{{\text{d}},\text{c}}^{\alpha},{\mathcal{R}}_{{\text{d}},\text{s}}^{\alpha})$ can be analyzed in a similar way we analyze $({\mathcal{R}}_{{\text{d}},\text{c}},{\mathcal{R}}_{{\text{d}},\text{s}})$. We find that ${\mathcal{R}}_{{\text{d}},\text{c}}^{\alpha}$ (or ${\mathcal{R}}_{{\text{d}},\text{s}}^{\alpha}$) achieves a smaller high-SNR slope than ${\mathcal{R}}_{{\text{d}},\text{c}}$ (or ${\mathcal{R}}_{{\text{d}},\text{s}}$), whereas ${\mathcal{R}}_{{\text{d}},\text{c}}^{\alpha}$ yields the same diversity order as ${\mathcal{R}}_{{\text{d}},\text{c}}$.

\section{Uplink Performance}
\subsection{Performance of Communications}
At the $l$th time slot of uplink ISAC, the BS receives
\begin{align}
{\textbf{y}}_{{\text{u}},l}=\sum\nolimits_{k=1}^{K}{\textbf{h}}_{{\text{u}},k}x_{k,l}+
{\textbf{G}}^{\mathsf{H}}{\textbf{s}}_l+{\textbf{n}}_{{\text{u}},l},
\end{align}
where ${\textbf{n}}_{{\text{u}},l}\sim{\mathcal{CN}}\left({\textbf{0}},{\textbf{I}}_N\right)$ denotes the $l$th column of ${\textbf{N}}^{\mathsf{H}}$. To approach the performance upper bound of communication signals, we use the minimum MSE (MMSE)-SIC decoder to detect the information bits, which is capacity-achieving \cite{Heath2018}. Moreover, from a worst-case design perspective \cite{Hassibi2003}, the aggregate interference-plus-noise ${\textbf{b}}_l={\textbf{G}}^{\mathsf{H}}{\textbf{s}}_l+{\textbf{n}}_{{\text{u}},l}\in{\mathbbmss{C}}^{N\times 1}$ is treated as the Gaussian noise. Accordingly, the uplink sum CR at the $l$th time slot is given by
\begin{align}
{\mathcal{R}}_{{\text{u}},l}=\log_2\det({\textbf{I}}_N+p_{\text{c}}{\textbf{H}}_{\text{u}}{\textbf{H}}_{\text{u}}^{\mathsf{H}}{\textbf{W}}_l^{-1}),
\end{align}
where ${\textbf{H}}_{\text{u}}=\left[{\textbf{h}}_{{\text{u}},1}\cdots{\textbf{h}}_{{\text{u}},K}\right]\in{\mathbbmss{C}}^{N\times K}$ and ${\textbf{W}}_l={\mathbbmss{E}}\left\{{\textbf{b}}_l{\textbf{b}}_l^{\mathsf{H}}\right\}=\varrho_l^2{\textbf{I}}_N$ with $\varrho_l^2=1+\left|{\textbf{s}}_l^{\mathsf{H}}{\textbf{R}}_{\text{T}}{\textbf{s}}_l\right|$. As a result, the uplink sum CR can be simplified to
${\mathcal{R}}_{{\text{u}},l}=\log_2\det({\textbf{I}}_N+{p_{\text{c}}}{\varrho_l^{-2}}{\textbf{H}}_{\text{u}}{\textbf{H}}_{\text{u}}^{\mathsf{H}})$.
It is worth mentioning that the uplink sum CR varies with the index of time slot, $l$. For brevity, we leverage the expectation of ${\mathcal{R}}_{{\text{u}},l}$ with respect to $l$ to evaluate the uplink performance of communication signals, namely ${\mathcal{R}}_{{\text{u}}}=\frac{1}{L}\sum_{l=1}^{L}{\mathcal{R}}_{{\text{u}},l}$.
\subsubsection{Outage Probability}
The OP of the uplink sum CR is given as $P_{\text{u}}=\Pr\left({\mathcal{R}}_{{\text{u}}}<{\mathcal{R}}\right)$. Using similar steps as those outlined in Appendix \ref{Proof_Theorem_Downlink_OP_Asymptotic}, we characterize the high-SNR behaviour of the OP as follows.
\vspace{-5pt}
\begin{theorem}\label{Theorem_Uplink_OP_Asymptotic}
In the high-SNR regime, the outage probability satisfies $\lim_{p_{\text{c}}\rightarrow\infty}P_{\text{u}}= {\mathcal{O}}\left(p_{\text{c}}^{-NK}\right)$.
\end{theorem}
\vspace{-5pt}
\vspace{-5pt}
\begin{remark}
A diversity order of $KN$ is achievable for the uplink sum communication rate.
\end{remark}
\vspace{-5pt}
\subsubsection{Ergodic Rate}
The uplink ECR of the CUs is given as ${\mathcal{R}}_{{\text{u}},{\text{c}}}={\mathbbmss{E}}\left\{{\mathcal{R}}_{{\text{u}}}\right\}$. Bearing the same idea built in Appendix \ref{Proof_Theorem_Downlink_ECR_Asymptotic} in mind, we obtain Theorem \ref{Theorem_Uplink_ECR_Asymptotic}.
\vspace{-5pt}
\begin{theorem}\label{Theorem_Uplink_ECR_Asymptotic}
The ECR satisfies $\lim_{p_{\text{c}}\rightarrow\infty}{\mathcal{R}}_{{\text{u}},{\text{c}}}= K\log_2{{p_{\text{c}}}}+\frac{1}{\ln{2}}\sum_{t=0}^{K-1}\left(
\sum_{a=1}^{M-t-1}\frac{1}{a}-{\emph{\textbf{C}}}\right)-\frac{K}{L}\sum_{l=1}^{L}\log_2{{\varrho_l^2}}$.
\end{theorem}
\vspace{-5pt}
\vspace{-5pt}
\begin{remark}
A high-SNR slope of $K$ is achievable for the uplink sum communication rate.
\end{remark}
\vspace{-5pt}

\subsection{Performance of Sensing}
After decoding all the information bits sent by the CUs, the BS can remove $\sum\nolimits_{k=1}^{K}{{\textbf{h}}}_{{\text{u}},k}{\textbf{x}}_{{\text{u}},k}^{\mathsf{H}}$ from ${\textbf{Y}}_{{\text{u}}}$ in \eqref{Uplink_Basic_Model}. The rest part can be used for radar sensing \cite{Chiriyath2016}, which is expressed as ${\textbf{Y}}_{\text{s}}={\textbf{G}}^{\mathsf{H}}{\textbf{S}}+{\textbf{N}}^{\mathsf{H}}$. Following similar steps as those outlined in Appendix \ref{Proof_Lemma_Downlink_Sensing_MI}, we can get the maximal uplink SR as follows
\begin{align}
{\mathcal{R}}_{{\text{u}},{\text{s}}}={N}{L^{-1}}\max\nolimits_{{\mathsf{tr}}\left({\textbf{S}}{\textbf{S}}^{\mathsf{H}}\right)\leq p_{\text{s}}}\log_2\det\left({\textbf{I}}_L+{\textbf{S}}^{\mathsf{H}}{\textbf{R}}_{\text{T}}{\textbf{S}}\right).
\end{align}
By the method we derive Theorem \ref{Theorem_Downlink_SR}, we obtain Theorem \ref{Theorem_SR_ER}.
\vspace{-5pt}
\begin{theorem}\label{Theorem_SR_ER}
The maximal uplink SR is given as ${\mathcal{R}}_{{\text{u}},{\text{s}}}={N}{L^{-1}}\sum\nolimits_{m=1}^{M}\log_2\left(1+\lambda_ma_m^{\star}\right)$, where $a_{m}^{\star}\!=\!\max\left\{0,\frac{1}{\nu}\!-\!\frac{1}{\lambda_m}\right\}$ with $\sum_{m=1}^{M}\!\max\left\{\!0,\frac{1}{\nu}\!-\!\frac{1}{\lambda_m}\right\}\!=\!p_{\text{s}}$. The SR is maximized when ${\textbf{S}}{\textbf{S}}^{\mathsf{H}}\!=\!{\textbf{U}}_{\text{T}}^{\mathsf{H}}{\bm\Theta}^{\star}{\textbf{U}}_{\text{T}}$ with ${\bm\Theta}^{\star}\!=\!{\mathsf{diag}}\left\{a_1^{\star},\cdots,a_M^{\star}\right\}$. When $p_{\text{s}}\rightarrow\infty$, we can obtain
\begin{align}
{\mathcal{R}}_{{\text{u}},{\text{s}}}
\approx\frac{NM}{L}\left(\log_2{p_{\text{s}}}+\frac{1}{M}\sum\nolimits_{m=1}^{M}\log_2\left(\frac{\lambda_m}{M}\right)\right).
\end{align}
\end{theorem}
\vspace{-5pt}
\vspace{-5pt}
\begin{remark}
The uplink SR achieves the same high-SNR slope, namely $\frac{NM}{L}$, as the downlink SR.
\end{remark}
\vspace{-5pt}

\subsection{Performance of FDSAC}\label{Section4c}
Turn now to the uplink FDSAC (U-FDSAC) system where the MMSE-SIC decoding and the optimal radar waveforming are exploited. The SR and the ECR are given by ${\mathcal{R}}_{{\text{u}},\text{s}}^{\alpha}=\frac{N(1-\alpha)}{L}\max_{{\mathsf{tr}}\left({\textbf{S}}{\textbf{S}}^{\mathsf{H}}\right)\leq p_{\text{s}}}\log_2\det\left({\textbf{I}}_L+\frac{1}{1-\alpha}{\textbf{S}}^{\mathsf{H}}{\textbf{R}}_{\text{T}}{\textbf{S}}\right)$ and ${\mathcal{R}}_{{\text{u}},\text{c}}^{\alpha}={\mathbbmss{E}}\left\{\alpha\log_2\det\left({\textbf{I}}_N+\frac{p_{\text{c}}}{\alpha}{\textbf{H}}_{\text{u}}{\textbf{H}}_{\text{u}}^{\mathsf{H}}\right)\right\}$ , respectively. Note that $({\mathcal{R}}_{{\text{u}},\text{c}}^{\alpha},{\mathcal{R}}_{{\text{u}},\text{s}}^{\alpha})$ can be discussed in the way we discuss $({\mathcal{R}}_{{\text{u}},\text{c}},{\mathcal{R}}_{{\text{u}},\text{s}})$. In particular, we find the high-SNR slope of ${\mathcal{R}}_{{\text{u}},\text{c}}^{\alpha}$ (or ${\mathcal{R}}_{{\text{u}},\text{s}}^{\alpha}$) is no larger than that of ${\mathcal{R}}_{{\text{u}},\text{s}}$ (or ${\mathcal{R}}_{{\text{u}},\text{c}}$). Moreover, we note that ${\mathcal{R}}_{{\text{u}},\text{c}}$ yields the same diversity order as ${\mathcal{R}}_{{\text{u}},\text{c}}^{\alpha}$.
\vspace{-5pt}
\begin{remark}
The results in Section \ref{Section3c} and Section \ref{Section4c} demonstrate that the ISAC system achieves a larger high-SNR slope than the FDSAC system in terms of both the CR and the SR. In other words, ISAC can provide more degrees of freedom \cite{Heath2018} for both the CR and the SR than FDSAC.
\end{remark}
\vspace{-5pt}

\section{Rate Region Characterization}
We now characterize the communication-sensing rate region of the considered ISAC and FDSAC systems. As stated before, we assume $p_{\text{c}}$ and $p_{\text{s}}$ are fixed values and do not consider the influence of power allocation. In light of this and in order to present the result with more generality, we now consider another case that $p_c$ and $p_{\text{s}}$ are subject to the constraints $p_{\text{c}}\in\left[0,\dot{p}_{\text{c}}\right]$ and $p_{\text{s}}\in\left[0,\dot{p}_{\text{s}}\right]$, where $\dot{p}_{\text{c}}$ and $\dot{p}_{\text{s}}$ denote the maximal communication SNR and the maximal sensing SNR, respectively. Let ${\mathcal{R}}_{\text{c}}$ and ${\mathcal{R}}_{\text{s}}$ denote the achievable ECR and SR, respectively. Based on our previous discussions, for ISAC systems, the achievable downlink rate region satisfies
\begin{equation}\label{Downlink_ISAC_RG}
\left\{\!\left({\mathcal{R}}_{\text{c}},\!{\mathcal{R}}_{\text{s}}\right)\!|{\mathcal{R}}_{\text{c}}\!\!\in\![0,\!{\mathcal{R}}_{{\text{d}},\text{c}}],\!
{\mathcal{R}}_{\text{s}}\!\in\![0,\!{\mathcal{R}}_{{\text{d}},\text{s}}],\!p_{\text{c}}\!\in\![0,\!\dot{p}_{\text{c}}],\!p_{\text{s}}\!=\!\dot{p}_{\text{s}}\!\right\},
\end{equation}
whereas the achievable uplink rate region satisfies
\begin{equation}\label{Uplink_ISAC_RG}
\left\{\!\left({\mathcal{R}}_{\text{c}},\!{\mathcal{R}}_{\text{s}}\right)\!|{\mathcal{R}}_{\text{c}}\!\!\in\![0,\!{\mathcal{R}}_{{\text{u}},\text{c}}],\!
{\mathcal{R}}_{\text{s}}\!\!\in\![0,\!{\mathcal{R}}_{{\text{u}},\text{s}}],\!p_{\text{s}}\!\!\in\![0,\!\dot{p}_{\text{s}}],p_{\text{c}}\!=\!\dot{p}_{\text{c}}\!\right\}.
\end{equation}
As for FDSAC systems, the downlink rate region satisfies
\begin{equation}\label{Downlink_FDSAC_RG}
\left\{\left({\mathcal{R}}_{\text{c}},{\mathcal{R}}_{\text{s}}\right)\left|
\begin{aligned}
&{\mathcal{R}}_{\text{c}}\in[0,{\mathcal{R}}_{{\text{d}},\text{c}}^{\alpha}]
,{\mathcal{R}}_{\text{s}}\in[0,{\mathcal{R}}_{{\text{d}},\text{s}}^{\alpha}] \\
&\alpha\in[0,1],p_{\text{c}}=\dot{p}_{\text{c}},
p_{\text{s}}=\dot{p}_{\text{s}}
\end{aligned}
\right.\right\},
\end{equation}
whereas the uplink rate region satisfies
\begin{equation}\label{Upink_FDSAC_RG}
\left\{\left({\mathcal{R}}_{\text{c}},{\mathcal{R}}_{\text{s}}\right)\left|
\begin{aligned}
&{\mathcal{R}}_{\text{c}}\in[0,{\mathcal{R}}_{{\text{u}},\text{c}}^{\alpha}]
,{\mathcal{R}}_{\text{s}}\in[0,{\mathcal{R}}_{{\text{u}},\text{s}}^{\alpha}] \\
&\alpha\in[0,1],p_{\text{c}}=\dot{p}_{\text{c}},
p_{\text{s}}=\dot{p}_{\text{s}}
\end{aligned}
\right.\right\}.
\end{equation}
\section{Numerical Results}
Numerical analysis is presented to evaluate the performance of ISAC systems. The parameters used for simulation are listed as follows: $N=2$, $M=2$, $L=4$, and $K=2$. The $(i,j)$th element of ${\textbf{R}}_{\text{T}}$ is set as $0.7^{|i-j|}$, whereas the $(i,j)$th element of ${\textbf{R}}_k={\textbf{R}}$ ($\forall k$) is set as $0.8^{|i-j|}$.

\begin{figure}[!t]
    \centering
    \subfigbottomskip=0pt
	\subfigcapskip=-5pt
\setlength{\abovecaptionskip}{0pt}
    \subfigure[Outage probability.]
    {
        \includegraphics[height=0.15\textwidth]{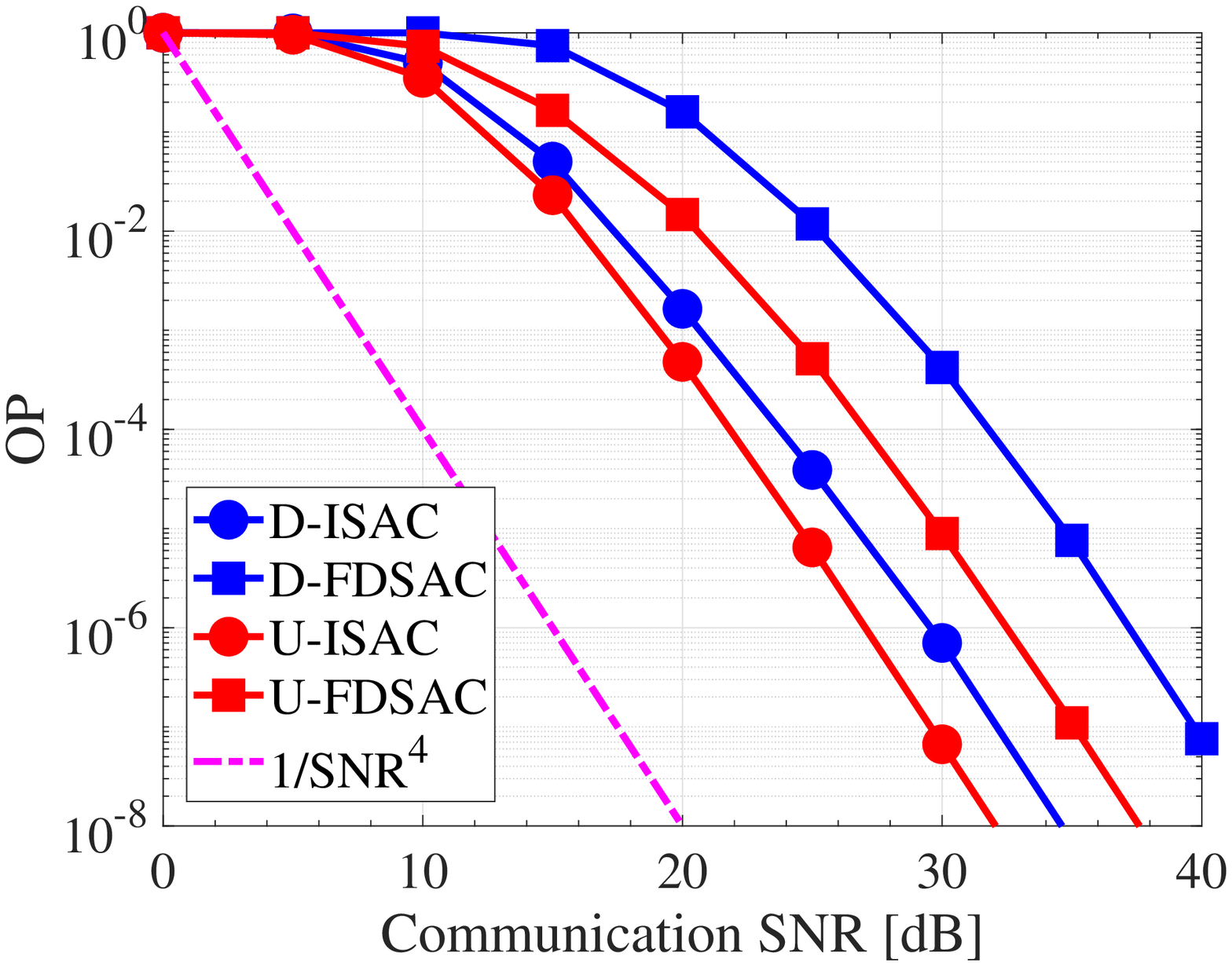}
	   \label{fig1a}	
    }
   \subfigure[Ergodic communication rate.]
    {
        \includegraphics[height=0.15\textwidth]{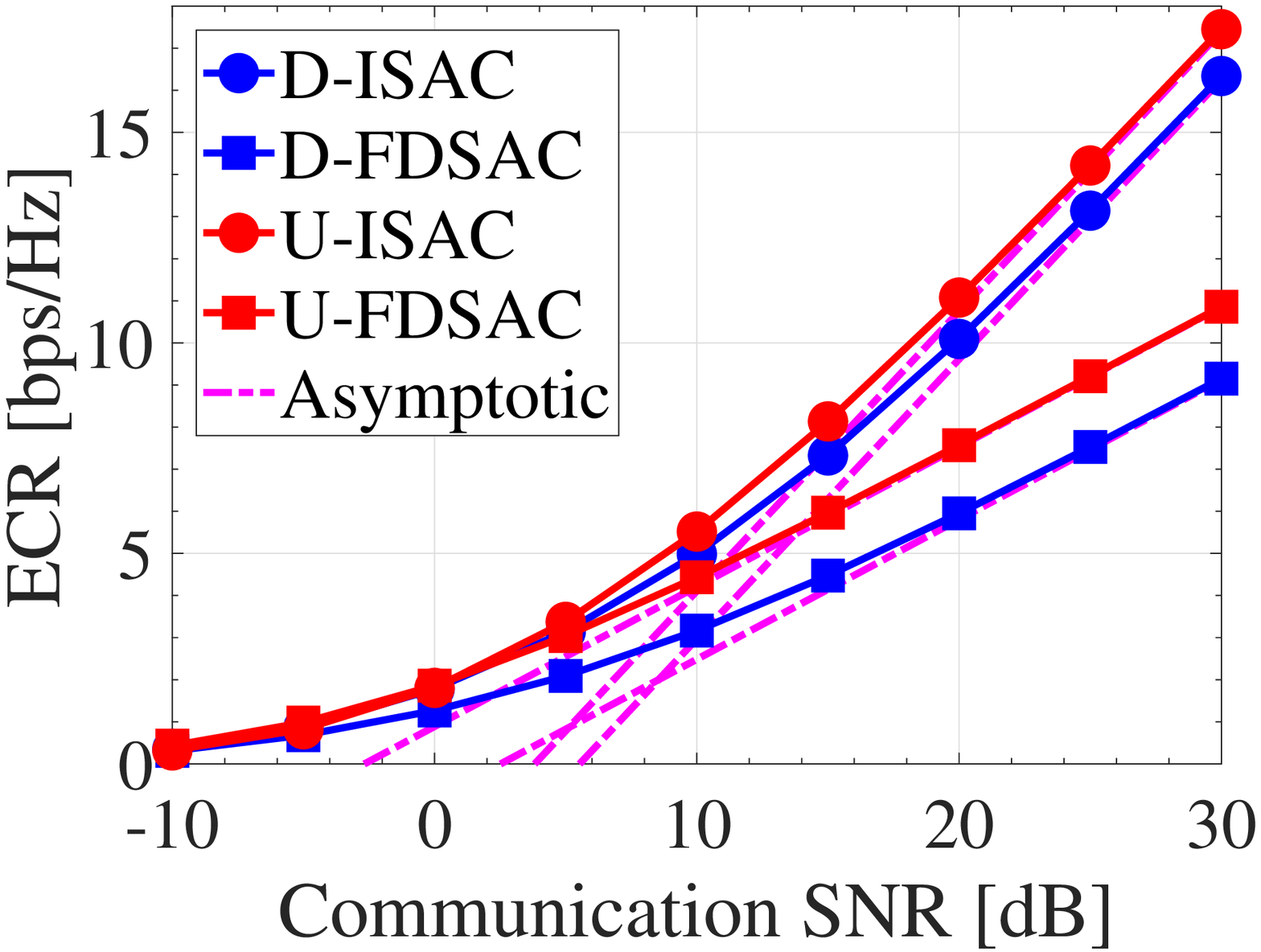}
	   \label{fig1b}	
    }
\caption{Performance of communications. ${\mathcal{R}}=5$ bps/Hz, $\alpha=0.5$, and $p_{\text{s}}=10$ dB.}
    \label{figure1}
\end{figure}

{\figurename} {\ref{fig1a}} and {\figurename} {\ref{fig1b}} plot the OP and the ECR versus the communication SNR, $p_{\text{c}}$, respectively. As shown in {\figurename} {\ref{fig1a}}, ISAC achieves a lower OP than FDSAC for both downlink and uplink transmissions. Besides, in the high-SNR regime, the OP curves for ISAC and FDSAC are mutually parallel, which suggests that these two systems achieve the same diversity order. This observation agrees with the conclusions drawn in Section \ref{Section3c} and Section \ref{Section4c}. Note that the curve for $p_{\text{c}}^{-4}$ is also provided to demonstrate the achievable diversity order. As shown, in the high-SNR regime, the curves for OP are parallel to the one for $p_{\text{c}}^{-4}$, suggesting the achievable diversity order obtained in the previous section is tight. Turn now to {\figurename} {\ref{fig1b}}. It can be observed that in the low-SNR regime, ISAC achieves virtually the same ECR as FDSAC, whereas in the high-SNR regime, ISAC achieves a higher ECR than FDSAC. The reason lies in that the ECR of ISAC systems yields a larger high-SNR slope than the ECR of FDSAC systems.

\begin{figure}[!t]
\centering
    \subfigbottomskip=0pt
	\subfigcapskip=-5pt
\setlength{\abovecaptionskip}{0pt}
    \subfigure[Downlink.]
    {
        \includegraphics[height=0.15\textwidth]{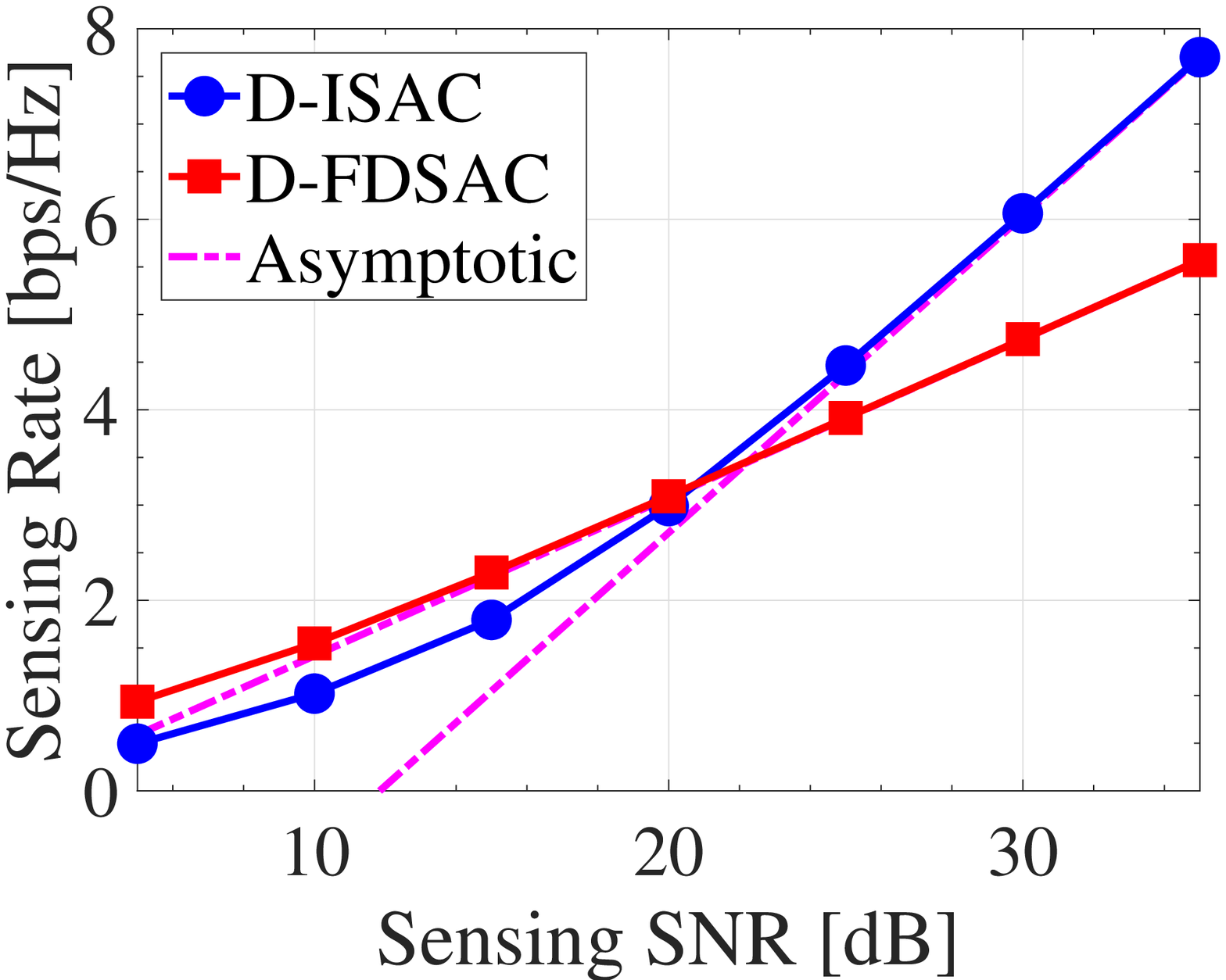}
	   \label{fig2a}	
    }
   \subfigure[Uplink.]
    {
        \includegraphics[height=0.15\textwidth]{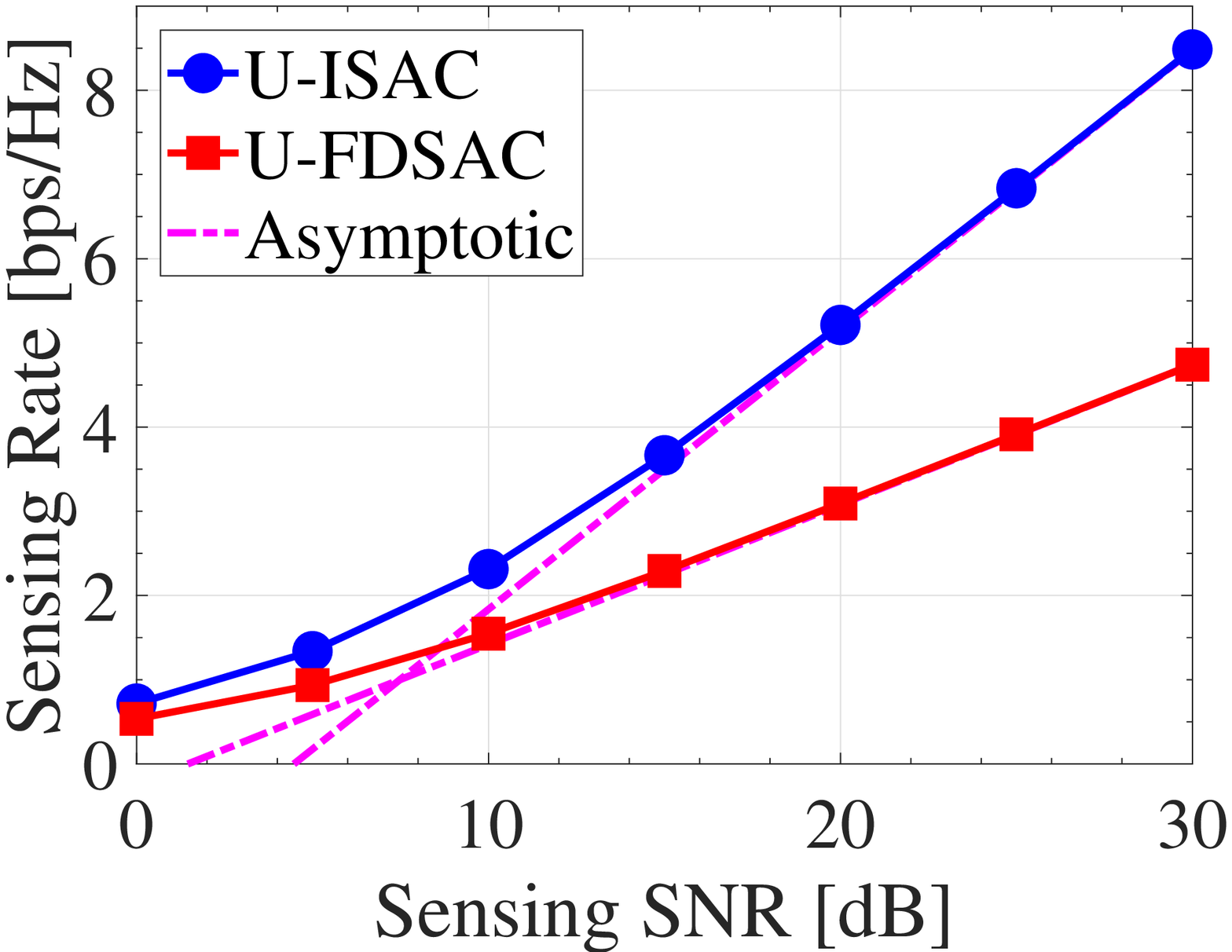}
	   \label{fig2b}	
    }
\caption{Performance of sensing. $\alpha=0.5$ and $p_{\text{c}}=5$ dB.}
\label{figure2}
\end{figure}

{\figurename} {\ref{fig2a}} and {\figurename} {\ref{fig2b}} plot the downlink and the uplink SRs versus the sensing SNR, $p_{\text{s}}$, respectively. As shown, the SR curve of ISAC yields a larger high-SNR slope than the SR curve of FDSAC. Besides, as {\figurename} {\ref{fig2a}} shows, in low and moderate SNR regions, D-FDSAC achieves a higher SR than D-ISAC. This is because the communication signal interferes in D-ISAC's sensing procedure, thus reducing the SR. Yet, involving a larger high-SNR slope, the SR of D-ISAC will exceed that of D-FDSAC as $p_{\text{s}}$ increases, which agrees with the observation from {\figurename} {\ref{fig2a}}. By contrast, as shown in {\figurename} {\ref{fig2b}}, the SR of U-ISAC is higher than that of U-FDSAC in regions of all SNR. This is because the communication signal has no influence on U-ISAC's sensing procedure under the SIC-based framework \cite{Chiriyath2016}.

\begin{figure}[!t]
    \centering
    \subfigbottomskip=0pt
	\subfigcapskip=-5pt
\setlength{\abovecaptionskip}{0pt}
    \subfigure[Downlink.]
    {
        \includegraphics[height=0.15\textwidth]{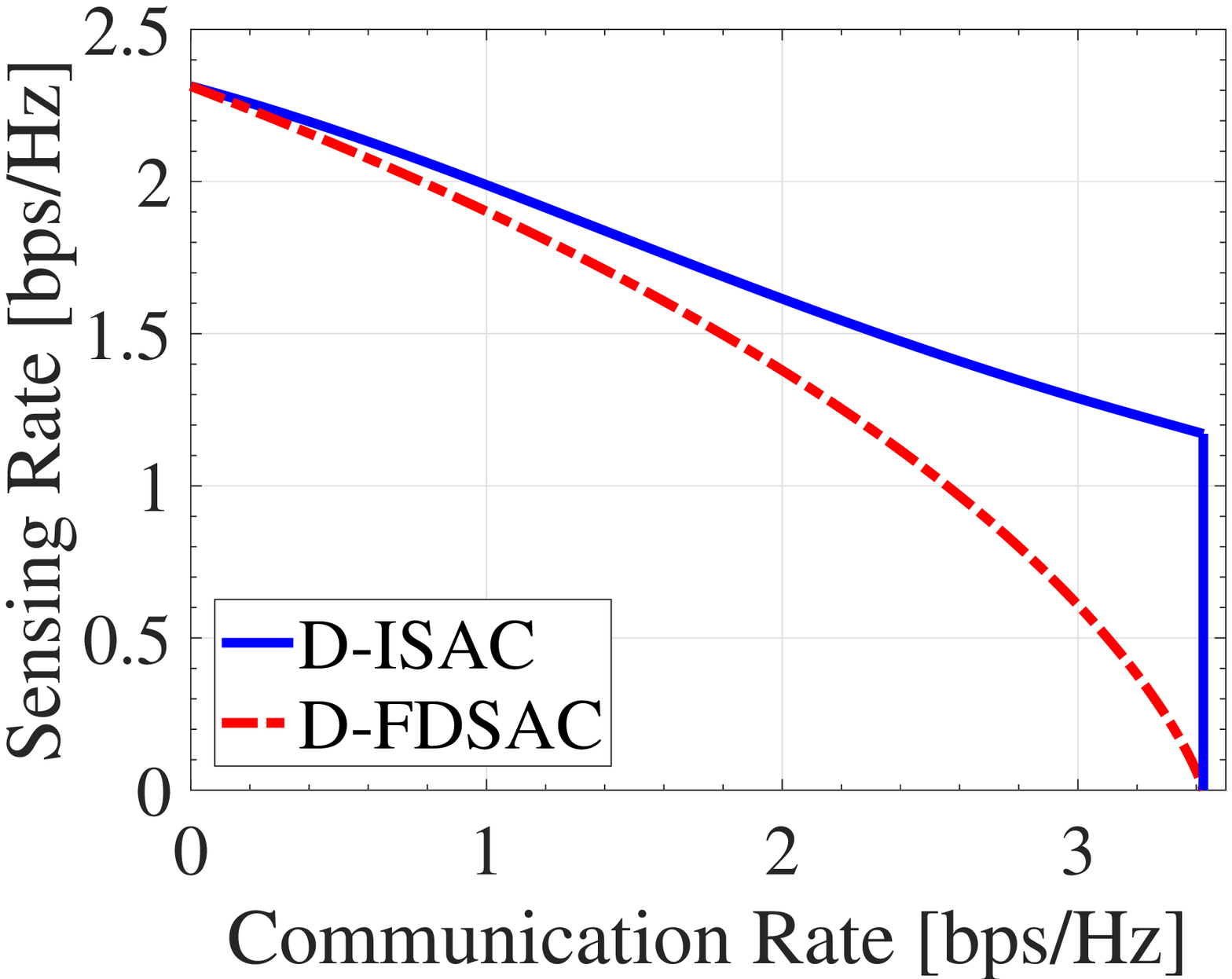}
	   \label{fig3a}	
    }
   \subfigure[Uplink.]
    {
        \includegraphics[height=0.15\textwidth]{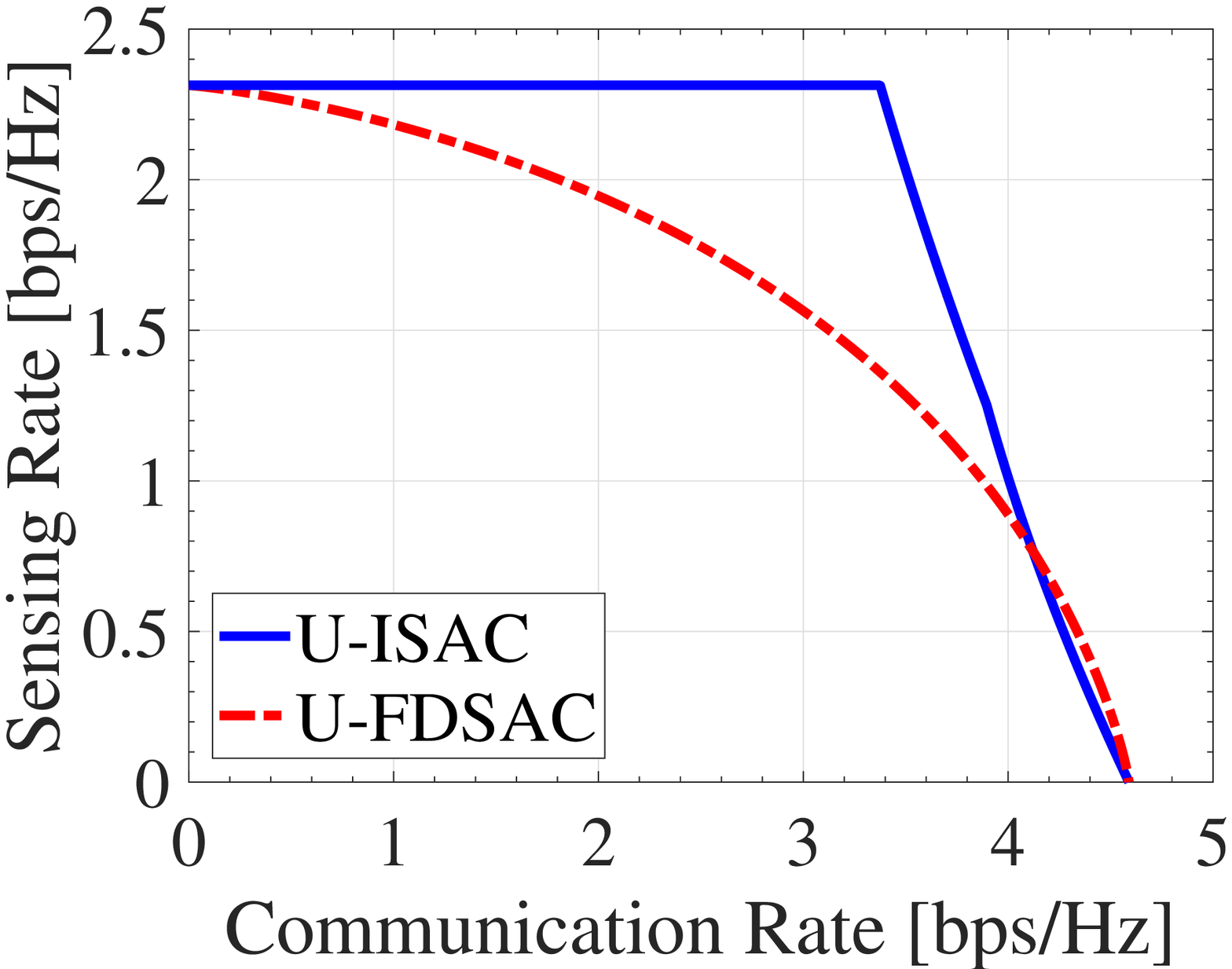}
	   \label{fig3b}	
    }
\caption{Rate region. $\dot{p}_{\text{c}}=5$ dB and $\dot{p}_{\text{s}}=10$ dB.}
    \label{figure3}
\end{figure}

Notably, it is challenging to provide a rigorous comparison of the CR-SR regions achieved by ISAC and FDSAC. As a compromise, we provide some numerical results in {\figurename} {\ref{figure3}} for a heuristic exploration. In particular, {\figurename} {\ref{fig3a}} compares the downlink rate regions of ISAC (presented in \eqref{Downlink_ISAC_RG}) and FDSAC (presented in \eqref{Downlink_FDSAC_RG}). As shown in this graph, the achievable rate region of D-FDSAC is entirely included in the achievable rate region of D-ISAC, which highlights the superiority of D-ISAC over conventional D-FDSAC. {\figurename} {\ref{fig3b}} compares the uplink rate regions of ISAC (presented in \eqref{Uplink_ISAC_RG}) and FDSAC (presented in \eqref{Upink_FDSAC_RG}). It can be observed that the rate region of U-FDSAC is mostly covered by that of U-ISAC. However, as {\figurename} {\ref{fig3b}} shows, in the high-ECR region, U-FDSAC yields a slightly higher SR than U-ISAC.

\section{Conclusion}
Communication and sensing performances of downlink and uplink ISAC systems have been analyzed. Closed-form analytical results have been developed to characterize the OP, ECR, SR, and CR-SR region, respectively. Simulation results have been provided to demonstrate the accuracy of the developed analytical results. Theoretical analyses have shown that ISAC can provide more degrees of freedom for both the CR and the SR than conventional FDSAC.

\begin{appendices}
\section{Proof of Theorem \ref{Theorem_Downlink_OP_Asymptotic}}\label{Proof_Theorem_Downlink_OP_Asymptotic}
Since $\lim_{p_{\text{c}}\rightarrow\infty}{\mathcal{R}}_{\text{d}}\!=\!\log_2\det({\textbf{I}}_K\!+\!\frac{p_{\text{c}}}{K}{\textbf{H}}_{\text{d}}^{\mathsf{H}}{\textbf{H}}_{\text{d}})$ \cite{Jindal2007}, we have $\lim_{p_{\text{c}}\rightarrow\infty}P_{\text{d}}\!=\!\lim_{p_{\text{c}}\rightarrow\infty}\Pr(\det({\textbf{I}}_K\!+\!\frac{p_{\text{c}}}{K}{\textbf{H}}_{\text{d}}^{\mathsf{H}}{\textbf{H}}_{\text{d}})<2^{{\mathcal{R}}})$. According to \cite{Yang2021}, when ${\textbf{R}}_k={\textbf{R}}$ ($\forall k\in{\mathcal{K}}$), we can get $\lim_{p_{\text{c}}\rightarrow\infty}P_{\text{d}}={\mathcal{O}}\left(p_{\text{c}}^{-MK}\right)$. Thus, the theorem is proved.
\section{Proof of Theorem \ref{Theorem_Downlink_ECR_Asymptotic}}\label{Proof_Theorem_Downlink_ECR_Asymptotic}
Since $\lim_{p_{\text{c}}\rightarrow\infty}{\mathcal{R}}_{\text{d}}\!=\!\log_2\det({\textbf{I}}_K+\frac{p_{\text{c}}}{K}{\textbf{H}}_{\text{d}}^{\mathsf{H}}{\textbf{H}}_{\text{d}})$ \cite{Jindal2007}, we have $\lim_{p_{\text{c}}\rightarrow\infty}{\mathcal{R}}_{{\text{c}},\text{d}}\!=\!\lim_{p_{\text{c}}\rightarrow\infty}{\mathbbmss{E}}\{\log_2\det(\frac{p_{\text{c}}}{K}{\textbf{H}}_{\text{d}}^{\mathsf{H}}{\textbf{H}}_{\text{d}})\}
=K\log_2{\frac{p_{\text{c}}}{K}}+\mathcal{E}_{\text{d}}$, where $\mathcal{E}_{\text{d}}\!=\!{\mathbbmss{E}}\{\log_2\det({\textbf{H}}_{\text{d}}^{\mathsf{H}}{\textbf{H}}_{\text{d}})\}$. When ${\textbf{R}}_k\!=\!{\textbf{I}}_{M}$ ($\forall k\in{\mathcal{K}}$), we can get $\mathcal{E}_{\text{d}}=\frac{1}{\ln{2}}\sum_{t=0}^{K-1}(
\sum_{a=1}^{M-t-1}\frac{1}{a}\!-\!{\emph{\textbf{C}}})$ with the aid of \cite{Heath2018}. Thus, the theorem is proved.
\section{Proof of Lemma \ref{Lemma_Downlink_Sensing_MI}}\label{Proof_Lemma_Downlink_Sensing_MI}
Denote ${\textbf{Z}}={\textbf{G}}^{\mathsf{H}}{\textbf{X}}_{\text{d}}+{\textbf{N}}=\left[{\textbf{z}}_1\cdots{\textbf{z}}_N\right]^{\mathsf{H}}$, where ${\textbf{z}}_n^{\mathsf{H}}={\textbf{g}}_n^{\mathsf{H}}{\textbf{X}}_{\text{d}}+{\textbf{n}}_n^{\mathsf{H}}$. It is worth noting that the row vectors of ${\textbf{Z}}$ are mutually independent, which satisfy ${\mathbbmss{E}}\{{\textbf{z}}_n{\textbf{z}}_n^{\mathsf{H}}\}
={\mathbbmss{E}}\{{\textbf{X}}_{\text{d}}^{\mathsf{H}}{\textbf{R}}_{\text{T}}{\textbf{X}}_{\text{d}}\}+{\textbf{I}}_L$ ($\forall n\in{\mathcal{N}}$). As stated before, we have ${\textbf{X}}_{\text{d}}=\left[{\textbf{x}}_{{\text{d}},1} \cdots {\textbf{x}}_{{\text{d}},L}\right]$, where ${\mathbbmss{E}}\{{\textbf{x}}_{{\text{d}},l}{\textbf{x}}_{{\text{d}},l'}^{\mathsf{H}}\}={\textbf{0}}$ ($\forall l\neq l'$) and ${\mathbbmss{E}}\{{\textbf{x}}_{{\text{d}},l}{\textbf{x}}_{{\text{d}},l}^{\mathsf{H}}\}={\bm\Sigma}_{{\textbf{H}}_{\text{d}}}$ ($\forall l\in\mathcal{L}$). It follows that ${\mathbbmss{E}}\{{\textbf{X}}_{\text{d}}^{\mathsf{H}}{\textbf{R}}_{\text{T}}{\textbf{X}}_{\text{d}}\}
={\mathbbmss{E}}\{{\mathsf{tr}}\left({\textbf{R}}_{\text{T}}{\bm\Sigma}_{{\textbf{H}}_{\text{d}}}\right){\textbf{I}}_L\}
={\mathsf{tr}}\left({\textbf{R}}_{\text{T}}{\bm\Sigma}\right){\textbf{I}}_L$ and thus ${\mathbbmss{E}}\{{\textbf{z}}_n{\textbf{z}}_n^{\mathsf{H}}\}=\sigma^2{\textbf{I}}_L$. Denote ${\textbf{Y}}_{{\text{d}}}^{\mathsf{H}}=\left[{\textbf{y}}_{{\text{d}},1}\cdots{\textbf{y}}_{{\text{d}},N}\right]\in{\mathbbmss{C}}^{L\times N}$, where ${\textbf{y}}_{{\text{d}},n}^{\mathsf{H}}={\textbf{g}}_n^{\mathsf{H}}{\textbf{S}}+{\textbf{z}}_n^{\mathsf{H}}$. When ${\textbf{z}}_n$ is treated as Gaussian noise following ${\mathcal{CN}}({\textbf{0}},\sigma^2{\textbf{I}}_L)$, we have ${\textbf{y}}_{{\text{d}},n}\sim{\mathcal{CN}}({\textbf{0}},{\textbf{S}}^{\mathsf{H}}{\textbf{R}}_{\text{T}}{\textbf{S}}+\sigma^2{\textbf{I}}_L)$. Let $h(\textbf{x})$ denote the entropy of the random variable $\textbf{x}$. It is worth noting that the columns of ${\textbf{Y}}_{{\text{d}}}^{\mathsf{H}}$ are independent and identically distributed, and thus $I\left({\textbf{Y}}_{\text{d}};{\textbf{G}}|{\textbf{S}}\right)=N(h({\textbf{y}}_{{\text{d}},n})-h({\textbf{z}}_n))$. With the aid of \cite{Tang2019,Heath2018}, we can get ${\mathcal{I}}_{{\text{d}},L}$ in Lemma \ref{Lemma_Downlink_Sensing_MI}.
\section{Proof of Theorem \ref{Theorem_Downlink_SR}}\label{Proof_Theorem_Downlink_SR}
Note that $\log_2\det({\textbf{I}}_L+\frac{1}{\sigma^{2}}{\textbf{S}}^{\mathsf{H}}{\textbf{R}}_{\text{T}}{\textbf{S}})$ equals the MI of a virtual MIMO channel $\dot{\textbf{y}}={\textbf{R}}_{\text{T}}^{1/2}\dot{\textbf{x}}+\dot{\textbf{n}}$ with ${\mathbbmss{E}}\{\dot{\textbf{x}}{\dot{\textbf{x}}}^{\mathsf{H}}\}={\textbf{S}}{\textbf{S}}^{\mathsf{H}}$ and $\dot{\textbf{n}}\sim{\mathcal{CN}}({\textbf{0}},\sigma^2{\textbf{I}}_M)$. Thus, when this MI is maximized, the eigenvectors of ${\textbf{S}}{\textbf{S}}^{\mathsf{H}}$ should equal the left eigenvectors of ${\textbf{R}}_{\text{T}}^{1/2}$, with the eigenvalues chosen by the water-filling procedure \cite{Heath2018}, which yields ${\mathcal{R}}_{{\text{d}},{\text{s}}}=\frac{N}{L}\sum\nolimits_{m=1}^{M}\log_2\left(1+\frac{1}{\sigma^{2}}\lambda_ms_m^{\star}\right)$. Here, $\left\{\lambda_m\right\}_{m=1}^{M}$ denote eigenvalues of ${\textbf{R}}_{\text{T}}$ and $s_{m}^{\star}=\max\left\{0,\frac{1}{\nu}-\frac{\sigma^2}{\lambda_m}\right\}$ with $\sum_{m=1}^{M}\max\left\{0,\frac{1}{\nu}-\frac{\sigma^2}{\lambda_m}\right\}=p_{\text{s}}$. When $p_{\text{s}}\rightarrow\infty$, we have $\nu\rightarrow0$ and thus $\sum_{m=1}^{M}\max\left\{0,\frac{1}{\nu}-\frac{\sigma^2}{\lambda_m}\right\}=\frac{M}{\nu}-\sum_{m=1}^{M}\frac{\sigma^2}{\lambda_m}=p_{\text{s}}$. Hence, $\lim_{p_{\text{s}}\rightarrow\infty}{\mathcal{R}}_{{\text{d}},{\text{s}}}=\frac{N}{L}\sum_{m=1}^{M}\log_2\left(\frac{\lambda_m}{M\sigma^2}\right)+\frac{NM}{L}\log_2\left(p_{\text{s}}+\sum_{m=1}^{M}\frac{\sigma^2}{\lambda_m}\right)$. Thus, the theorem is proved.
\end{appendices}

\end{document}